# Uncertainty-Driven Modeling of Microporosity and Permeability in Clastic Reservoirs Using Random Forest

Muhammad Risha[1*], Mohamed Elsaadany[2], Paul Liu[1]

[1]North Carolina State University, Marine, Earth and Atmospheric Sciences Department, Raleigh, NC 27695, USA

[2]Geoscience Department, Universiti Teknologi PETRONAS, Seri Iskandar 32610, Perak, Malaysia

**Abstract**

Predicting microporosity and permeability in clastic reservoirs is a challenge in reservoir quality assessment, especially in formations where direct measurements are difficult or expensive. These reservoir properties are fundamental in determining a reservoir's capacity for fluid storage and transmission, yet conventional methods for evaluating them, such as Mercury Injection Capillary Pressure (MICP) and Scanning Electron Microscopy (SEM), are resource-intensive. The aim of this study is to develop a cost-effective machine learning model to predict complex reservoir properties using readily available field data and basic laboratory analyses. A Random Forest classifier was employed, utilizing key geological parameters such as porosity, grain size distribution, and spectral gamma-ray (SGR) measurements. An uncertainty analysis was applied to account for natural variability, expanding the dataset, and enhancing the model's robustness. The model achieved a high level of accuracy in predicting microporosity (93%) and permeability levels (88%). By using easily obtainable data, this model reduces the reliance on expensive laboratory methods, making it a valuable tool for early-stage exploration, especially in remote or offshore environments. The integration of machine learning with uncertainty analysis provides a reliable and cost-effective approach for evaluating key reservoir properties in siliciclastic formations. This model offers a practical solution to improve reservoir quality assessments, enabling more informed decision-making and optimizing exploration efforts.

[*]**Corresponding Author: mrisha@ncsu.edu**

# 1. Introduction

Accurately predicting microporosity and permeability in clastic reservoirs can be crucial for assessing reservoir quality, particularly in complex geological environments where these properties are difficult and costly to measure directly (Joseph et al., 2013; Van Geet et al., 2003). By utilizing machine learning models, these predictions can be made efficiently using readily available field data and inexpensive lab analyses, significantly improving decision-making in reservoir characterization. These parameters directly influence the capacity of a reservoir to store and transmit fluids, making them critical for hydrocarbon exploration and production (Nelson, 2011). The presence of clay minerals such as kaolinite, illite, and chlorite further complicates this evaluation, as they can either preserve or reduce porosity and permeability through various diagenetic processes (Kjølstad, 2014). Therefore, developing models that can predict these parameters under different geological conditions is a key focus of current reservoir studies (Alansari et al., 2019).

Machine learning models have become powerful tools for addressing this challenge, providing the ability to process large datasets and capture complex relationships between geological, petrophysical, and diagenetic variables (Tariq et al., 2021). By incorporating various inputs such as porosity, grain size distribution, and clay content, these models can predict the quality of reservoirs in clastic formations with a higher degree of accuracy (Ajdukiewicz & Lander, 2010). Specifically, Random Forest algorithms have proven effective due to their robustness in handling diverse data inputs and their ability to reduce overfitting, which is crucial when working with heterogeneous geological data (Xi et al., 2015). On top of that, Random Forest can handle missing values in predictor variables through imputation techniques, ensuring that incomplete datasets do not hinder predictive accuracy (Carranza & Laborte, 2016). This flexibility is vital in geological applications, where complete datasets are often difficult to obtain. The algorithm's strength in dealing with high-dimensional data and missing information makes it an ideal choice for predicting reservoir quality, even in data-limited scenarios (Rodriguez-Galiano et al., 2015).

However, geological data inherently carries uncertainty due to variability in measurements and sample heterogeneity. For example, microporosity and permeability can vary significantly across different facies, complicating efforts to model these properties consistently (Merletti et al., 2016; Worden & Burley, 2003) (Figure 1).

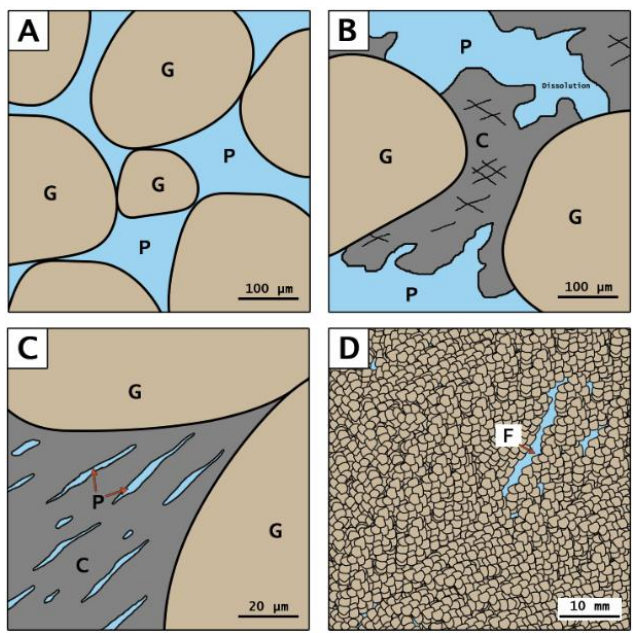

*Figure 1. Schematic representation of the main porosity types in clastic reservoirs. (A) Intergranular porosity consists of primary pore spaces between detrital grains, providing essential storage and flow pathways. (B) Dissolution porosity forms when carbonate cement is partially dissolved, creating additional pore spaces that enhance reservoir quality. (C) Microporosity occurs within clay-rich matrices, where small pores retain fluids but significantly reduce permeability. (D) Fracture porosity develops when fractures cut through a compacted grain framework, improving permeability in otherwise low-porosity rocks. (G – Grains, P – Porosity, C – Clay, F – Fracture).*

Addressing this variability is critical for creating models that not only predict the mean behavior of the reservoir but also account for the range of possible outcomes (Taylor et al., 2010). To achieve this, uncertainty analysis can be employed, allowing models to generate a broader set of predictions that better reflect the true range of geological conditions (Wellmann & Caumon, 2018; Yong et al., 2021).

The research process involved field data collection, sample preparation, and laboratory analyses, followed by the integration of petrographic and petrophysical data to understand microporosity and permeability in clastic reservoirs. While all stages were crucial to the overall study, the focus of this paper is on the machine learning model development for predicting microporosity and permeability, based on the collected data (Figure 2). One of the key advantages of this model is its ability to predict difficult-to-obtain properties, such as microporosity and permeability, using basic field information and inexpensive laboratory analyses. Traditionally, these properties are measured through methods like MICP and SEM, which are both expensive and time-consuming. For example, laboratory-based permeability measurements often require core sampling and specialized equipment, which can take weeks or months to process and incur significant costs (Zhong et al., 2019). Machine learning approaches offer a faster, more cost-effective alternative (Xu et al., 2022) by leveraging widely available data such as porosity, grain size distribution, and gamma-ray logs.

By integrating a robust uncertainty analysis, the model ensures that the variability in input data is adequately captured, further improving the reliability of predictions. This method not only reduces the need for expensive laboratory techniques but also enables the rapid prediction of reservoir properties from field data that can be collected more easily and affordably (Risha, 2024). Such an approach significantly reduces the cost and time required for reservoir characterization, making it particularly valuable in exploration projects where access to advanced laboratory resources may be limited (Arigbe et al., 2019).

The machine learning model developed in this paper has the potential to significantly reduce uncertainty in reservoir quality assessment, leading to better decision-making in hydrocarbon exploration.

## 2. Methodology

This study utilized field data collection, and laboratory analysis, used with machine learning techniques to predict microporosity and permeability in clastic reservoirs (Figure 2). The methodology involved gathering data from clastic outcrops on Labuan Island, Malaysia, which provided the key inputs for the machine learning model. The outcrops on Labuan Island, which form part of the larger Borneo geological structure, represent a range of formations, including the Crocker, Temburong, and Belait Formations, each characterized by unique depositional environments and diagenetic histories (Risha, 2025b). These formations reflect the regional tectonic and sedimentological processes of the Sabah Basin, making the island an ideal location for collecting analog data applicable to offshore hydrocarbon exploration (Madon, 1994; Nazaruddin et al., 2016).

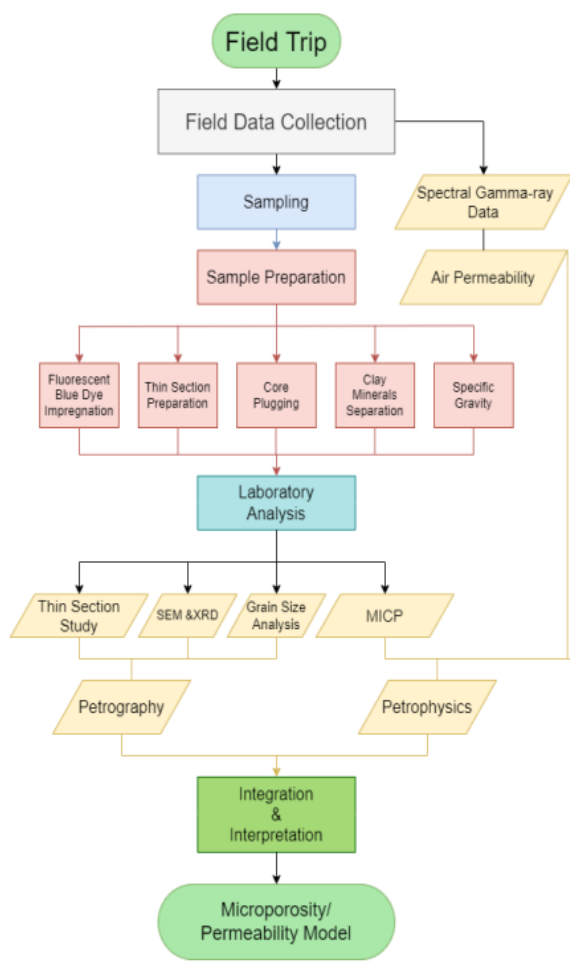

Figure 2. Workflow for data collection, preparation, and analysis leading to the development of the microporosity and permeability prediction model. The process includes field data collection, sample preparation, various laboratory analyses (e.g., MICP, SEM, XRD, and grain size analysis), and the integration of petrographic and petrophysical data to create a machine learning-based prediction model.

**2.1 Data Collection and Preparation**

Field data was collected from 16 outcrops across Labuan Island in Malaysia, which were chosen for their accessibility and representative nature of the region's stratigraphy (Risha & Douraghi, 2021). The island is located off the northwest coast of Borneo, forming part of the Brunei-Sabah Basin (Figure 3).

The geology of the island is mainly clastic formations that range from deep marine to shallow marine environments, with significant clay content that affects reservoir quality through diagenetic processes (Madon, 1997). The data collection process was carried out with the aim of measuring the fundamental petrophysical properties of the rocks, including thin section total porosity, air permeability, and grain size distribution, which are essential for predicting reservoir quality in clastic formations.

Porosity was measured using multiple techniques to ensure comprehensive analysis. Thin section analysis was conducted on rock samples to visually estimate porosity under a polarized microscope using blue impregnated epoxy which is helpful in identifying pore spaces. Additionally, Helium porosimetry was employed as well to measure porosity, providing a more accurate estimation of the interconnected pore spaces within the rock samples (Chastre & Ludovico-Marques, 2018). MICP analysis was another key method used to quantify microporosity (Risha, 2025a). It has provided critical insights into the pore sizes distribution, particularly in identifying pore sizes smaller than 2 μm in this research.

Permeability was measured using two approaches. First, air permeability was assessed in both horizontal and vertical orientations using the TinyPerm II Air-permeameter. This field-based method provided a quick, cost-effective and non-destructive estimation for permeability directly at the outcrop. Horizontal and vertical measurements were taken to account for anisotropy in permeability due to layering and sedimentary structures within the rock. Second, laboratory-based MICP analysis was conducted to obtain more detailed permeability data, particularly in tight rock samples where microporosity dominates (Taylor et al., 2010).

Clay, sand, and silt content were quantified using a combination of sieve analysis for the coarser fractions and hydrometer analysis for the finer particles as an assessment of the grain size distribution within each sample. SGR analysis was also performed on the outcrops to provide estimates of the clay content in terms of type, with specific attention to clay minerals like kaolinite, illite, and chlorite, which are known to influence porosity and permeability (Alansari et al., 2019). XRD and SEM were also used for few samples to identify the specific types of clay minerals and their spatial distribution within the pore spaces (Wilson & Pittman, 1977).

The data collected from these various field and laboratory methods formed the foundation for developing the machine learning model. Each measurement including porosity, permeability, or clay content was carefully chosen to reflect the primary factors influencing reservoir quality in clastic rocks. The integration of these diverse datasets ensures a comprehensive understanding of the factors controlling microporosity and permeability in the formations studied.

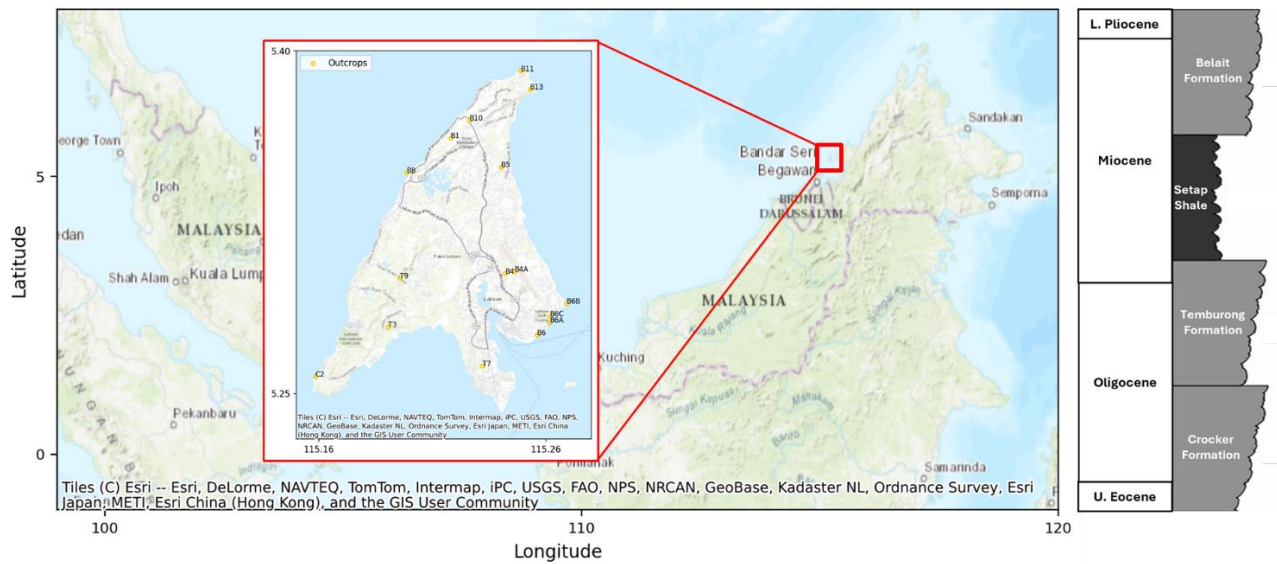

*Figure 3. Regional and geological context of the study area. The maps show study area, and selected outcrop locations across Labuan Island, highlighted within a regional context. The stratigraphic column illustrates the relative chronostratigraphic succession of geological formations on the island, ranging from the Upper Eocene to the Late Pliocene, providing insights into its depositional history (Hennig-Breitfeld et al., 2019; Risha et al., 2023).*

**2.2 Uncertainty Analysis**

To enhance the machine learning model's robustness and its ability to generalize similar geological conditions, an uncertainty analysis was applied. Model Generalization allows the model to perform better on unseen datasets. This process allowed for a broader representation of feature variability, ensuring the model could better capture the natural heterogeneity of clastic formations (Stracuzzi et al., 2018).

This uncertainty analysis played a crucial role in preparing the machine learning model to handle real-world variability. By training the model on a broader and more representative dataset, the analysis ensured that the model was better equipped to predict microporosity and permeability under similar range of geological conditions without the necessity of data from the same outcrops. Incorporating these uncertainty ranges minimized the risk of overfitting, enabling the model to produce more reliable and accurate predictions for reservoir quality (Janjuhah et al., 2019; Valdez et al., 2020).

Key input parameters subject to uncertainty included total porosity, air permeability, sand, silt, and clay content, the thorium-potassium (ThK) ratio, the sand-to-mud (SM) ratio, mean grain size, standard deviation, skewness, permeability, and microporosity (<2μm). These parameters were chosen due to their significant influence on reservoir quality and their role in controlling fluid flow and storage capacity.

Uncertainty ranges were applied to each parameter to simulate natural geological variability. Porosity was varied by ±4%, air permeability by ±10 millidarcies, and clay, sand, and silt contents by ±2% while the ThK ratio was adjusted by ±0.5, while the SM ratio, mean grain size, standard deviation and skewness we recalculated based on the new generated values. Target values (permeability and microporosity (<2μm)) were varied by ±10 millidarcies and ±5%, respectively. These ranges were based on observed field variability, ensuring realistic adjustments within the natural geological ranges (Table 1).

The study used 41 samples collected from the study area. Multiple field and lab measurements were taken to general a total of 308 measurements. The resampling process generated 200 new datasets, resulting in a final dataset of 61,600 data points. For each sample and measurement, random values we resampled within the specified uncertainty ranges of the original measurements. This process simulated the natural variability of geological formations as well as human error.

The resampled dataset thus provided a comprehensive representation of possible input combinations, improving the model's ability to generalize across different reservoir conditions.

*Table 1. Predefined Uncertainty Ranges in the Uncertainty resampling Model.*

| Parameter | Range | Unit | Remarks |
|---|---|---|---|
| Porosity | ±4% | Percentage (%) | Based on average variability |
| Air Permeability | ±10 | Millidarcies (mD) | Based on average variability |
| Clay Content | ±2% | Percentage (%) | **Adjusted to simulate error of sieve and hydrometer tests. Their total has to be 100%** |
| Sand Content | ±2% | Percentage (%) | |
| Silt Content | ±2% | Percentage (%) | |
| ThK Ratio | ±0.5 | Ratio | Based on average variability |
| SM Ratio | **Recalculated** | N/A | Recalculated from updated values |
| Mean Grain Size | **Recalculated** | Millimeters (mm) | Based on human estimation error |
| Standard Deviation | **Recalculated** | N/A | Based on human estimation error |
| Skewness | **Recalculated** | N/A | Based on human estimation error |
| Target Permeability | ±10 | Millidarcies (mD) | Based on average variability |
| Target Microporosity | ±5% | Percentage (%) | Based on average variability |

**2.3 Machine Learning Model Development**

This study employed a Random Forest classifier to predict microporosity and permeability levels in clastic reservoirs. Random Forest was selected because of its ability to handle complex, multi-dimensional geological data with both continuous and categorical variables (Risha & Liu, 2025). By aggregating the results of multiple decision trees, Random Forest reduces the likelihood of overfitting and enhances prediction accuracy, especially when dealing with noisy geological data (Zou et al., 2021).

The model utilized input variables that are crucial to reservoir quality assessment, including porosity, air permeability, clay content, grain size distribution, ThK ratio, sand-to-mud (SM) ratio, and the graphic sedimentary parameters. The model's target variables were permeability categories and microporosity domination (Figure 4). Permeability was classified into three categories: (Poor to Fair), (Moderate), and (Good to Very Good) (Table 2) (Risha, 2025b). This classification helps in distinguishing between different production potentials of the reservoir. Additionally, microporosity domination was used to classify samples as either macroporosity or microporosity-dominated, a critical distinction that affects the fluid flow within the reservoir. Porosity and permeability were key variables as they directly influence the reservoir's ability to store and transmit fluids. The ThK ratio is a proxy for clay content, and the grain size distribution, which includes sand, silt, and clay for the rock's texture.

*Table 2. Qualitative evaluation for permeability ranges*

| Permeability Range | Qualitative Description |
|---|---|
| Poor to Fair | <15 |
| Moderate | 16-50 |
| Good to V. Good | <50 |

The Random Forest model was configured with 100 decision trees, each limited to a maximum of five leaves. A decision tree works by splitting the dataset into subsets based on feature values, which can be thought of as a sequence of "if-then" rules. Each branch of the tree represents a decision, and the leaves are the final predictions. By limiting the maximum number of leaves to five, it was ensured that the model can avoid overfitting. With a large number of shallow trees (100 trees with 5 leaves each), the model captures the key relationships between input variables while avoiding learning noise or irrelevant patterns (Breiman, 2001; Salles et al., 2015).

The final training process used 90% of the dataset for training the Random Forest model, while the remaining 10% was reserved for evaluation. Training refers to the phase in which the model learns from the data by identifying patterns between the input features (such as porosity, permeability, and clay content) and the target variables (permeability and microporosity categories). By exposing the model to this portion of the data, it can "learn" relationships and develop the ability to make predictions on unseen data. Another testing phase was done by applying the model on a completely separate holdout dataset to evaluate how well the trained model performs on unseen data to check the model's ability to generalize to real-world cases.

To validate the model during the training phase and prevent overfitting, 5-fold cross-validation was applied. In this technique, the training data (90%) was divided into five equal parts, or (folds). The model was then trained on four of these folds and validated on the remaining fold. This process was repeated five times, with each fold serving as the validation set once. The average performance across all five iterations was recorded. Cross-validation not only enhances the model's reliability but also reduces the risk of overfitting, as each portion of the data is used for both training and validation at different stages (Gorriz et al., 2024; Kärkkäinen, 2014).

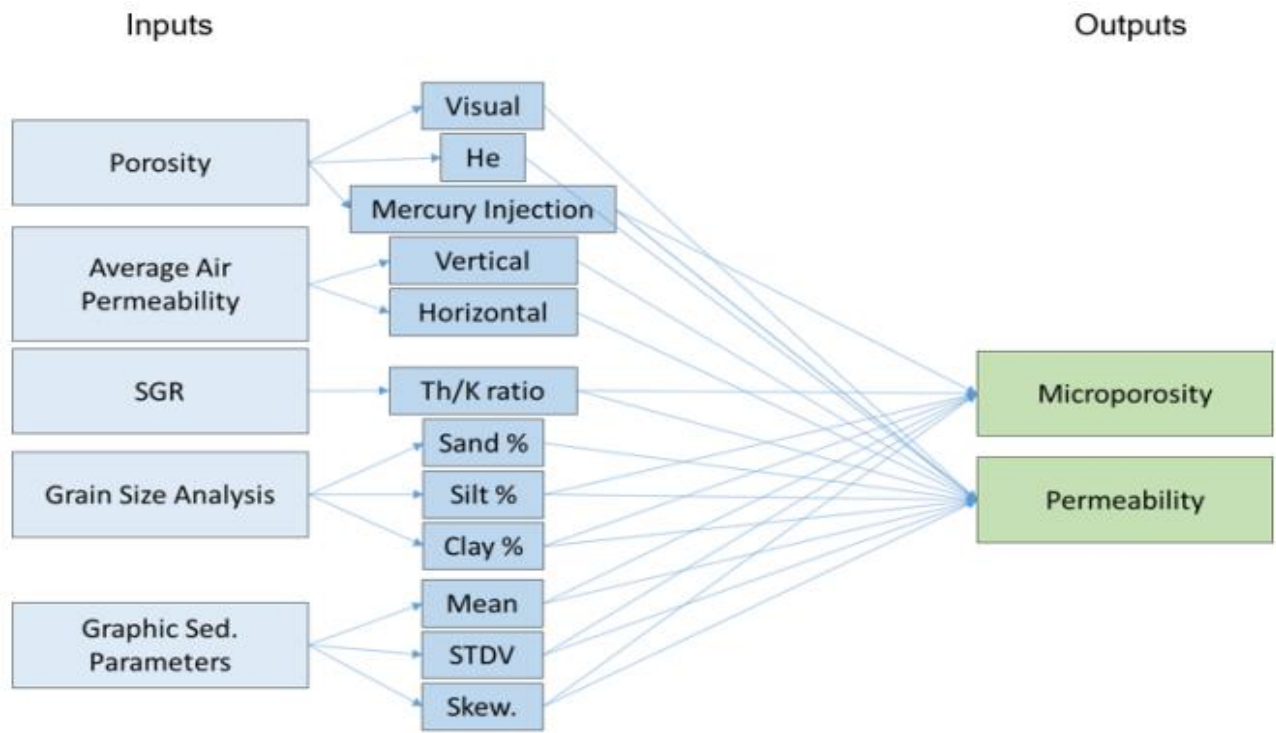

*Figure 4. A schematic representation of the Random Forest model inputs and outputs. The inputs consist of Porosity, Air Permeability, Th/K Ratio, Grain size data, and Graphic sedimentary parameters, while the outputs are descriptive permeability level and microporosity domination (<2μm).*

Before training the model, categorical variables, such as permeability and microporosity levels, were converted into numeric representations. This process, called encoding, was necessary because the machine learning algorithm operates on numerical data. Each categorical level was assigned a numerical code, allowing the Random Forest algorithm to process and learn from these features effectively. The encoding ensured that the model could properly interpret permeability and microporosity levels, which were crucial target variables for prediction (Smith et al., 2024).

To assess the performance of the Random Forest classifier, the Receiver Operating Characteristic (ROC) curve and Area Under the Curve (AUC) metrics were employed. The ROC curve provides a graphical representation of the model's diagnostic ability by plotting the True Positive Rate (TPR) against the False Positive Rate (FPR) at various threshold settings. Mathematically, the TPR (also known as sensitivity or recall) is calculated as the ratio of correctly predicted positive observations to the total actual positives, while the FPR is the ratio of incorrectly predicted positive observations to the total actual negatives (Walter, 2005). The AUC value, which ranges from 0 to 1, serves as a scalar measure summarizing the model's overall discriminatory ability, with values closer to 1 indicating the highest performance. In this study, the ROC curve was calculated for each class within the target variables using the predicted probabilities produced by the Random Forest model (Jadhav, 2020; Prati et al., 2011). These probabilities, rather than binary classifications, allowed for a more nuanced evaluation of the model's performance across a spectrum of thresholds. After completing the training and cross-validation steps, the holdout set was evaluated using the model to ensure an unbiased evaluation of the model's generalization capabilities.

## 3. Results

This section presents the results of the Random Forest model's performance in predicting permeability and microporosity domination, evaluated through cross-validation, an internal 10% evaluation set, and an independent holdout dataset. The model's accuracy was measured using confusion matrices, and the impact of uncertainty in the input parameters was also analyzed.

## 3.1 Model Accuracy

The Random Forest model was trained to predict both permeability categories and microporosity domination. The cross-validation error was 11.83% for permeability and 6.76% for microporosity, reflecting the model's strong performance during the training process. To further validate the model, it was tested on a 10% evaluation set, which was excluded from the training data. The evaluation accuracies on this dataset were 87.82% for permeability and 92.69% for microporosity domination. These high accuracy values indicate that the model was able to accurately capture the relationships between the input features and the target variables (Figure 5).

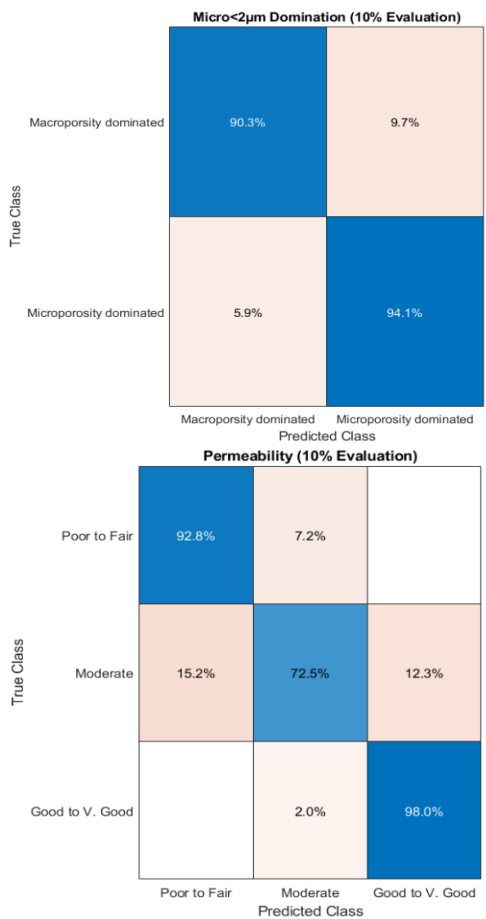

*Figure 5. Confusion matrices for permeability and microporosity predictions (10% Evaluation Set). (Top) The confusion matrix for microporosity indicates a high accuracy (94.1%) for the "Microporosity dominated" class, with some misclassification between "Macroporosity dominated" and "Microporosity dominated." (Bottom) The confusion matrix for permeability categories shows high accuracy in predicting "Poor to Fair" and "Good to Very Good" of (92.8%) and (98.0%) respectively, with performance in the "Moderate" category of (72.5%). Misclassifications mainly occurred between adjacent classes, such as "Poor to Fair" and "Moderate."*

## 3.2 Uncertainty Impact

Uncertainty in the input parameters had a measurable impact on the model's predictions. To assess the effect of incorporating uncertainty in the input data, the accuracy of predictions with and without uncertainty analysis was compared.

The model's accuracy for permeability and microporosity was higher when uncertainty was incorporated into the training process, as it allowed the model to capture the natural variability in the input features. This broader range of input values ensured that the model was not overly sensitive to small changes in the training data, resulting in more stable predictions on the evaluation and holdout datasets.

The uncertainty analysis had a significant impact on improving the model's accuracy by taking into account the natural variability present in geological data. By expanding the dataset to include this range of variability, the model was exposed to a broader spectrum of conditions during training. This approach helped mitigate the effects of measurement variability and reduced the likelihood of overfitting, enhancing the model's ability to generalize to new, unseen data. As a result, the prediction error decreased considerably, from 19.51% to 11.83% for permeability and from 9.87% to 6.76% for microporosity. These improvements highlight the critical role of uncertainty analysis in refining model performance and achieving more reliable predictions in complex geological contexts.

## 3.3 Holdout Set Evaluation

The model's generalization capability was further tested using an independent holdout dataset that was not used during training or internal evaluation. The evaluation accuracies on this independent holdout dataset were 90.91% for permeability and 81.82% for microporosity domination.

Permeability type showed perfect 100% accuracy predicting (Poor to Fair) and (Moderate) permeability, while misclassifying only one sample of (Good to V. Good) as (moderate) permeability calculated as 80% accuracy. For porosity type domination, the confusion matrix shows that the model correctly predicted 90.3% of the macroporosity-dominated samples, with 9.7% misclassified as microporosity-dominated. For the microporosity-dominated samples, 94.1% were correctly predicted, with 5.9% incorrectly predicted as macroporosity-dominated (Figure 6).

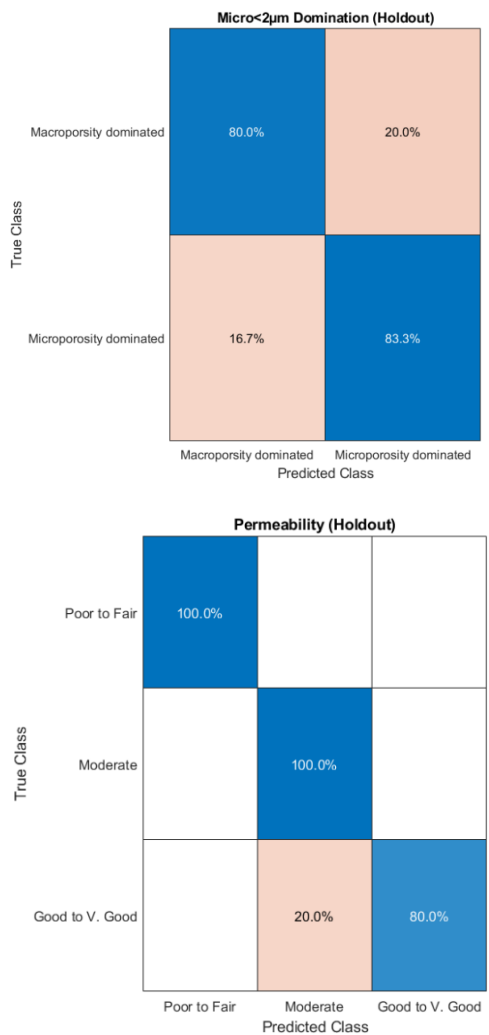

*Figure 6. Confusion matrices for permeability and microporosity predictions using the Random Forest model (Holdout Set). (Left) The confusion matrix for permeability shows 100% accuracy in predicting both "Poor to Fair" and "Moderate" permeability categories, with slight misclassification in the "Good to Very Good" category (80% accuracy). The model performed very well on the holdout set, demonstrating strong generalization. (Right) The confusion matrix for microporosity predictions displays good accuracy, with 83.3% of "Microporosity dominated" cases correctly classified, though there was some misclassification between "Macroporosity dominated" and "Microporosity dominated" classes (80% and 16.7% misclassification, respectively). The model's performance on the holdout set reflects its capability to generalize predictions to new, unseen data.*

## 4. Discussion

This section provides an interpretation of the model's results, focusing on the importance of the uncertainty analysis, the implications of the model's accuracy, and potential areas for improvement in future work.

**4.2 Model Accuracy and Implications**

The high accuracy of the model in predicting permeability and microporosity has significant implications for reservoir quality assessment. With cross-validation errors of 11.83% for permeability and 6.76% for microporosity, the model demonstrated strong predictive power. Evaluation on the holdout dataset resulted in accuracies of 90.91% for permeability and 81.82% for microporosity domination, confirming the model's generalization capabilities.

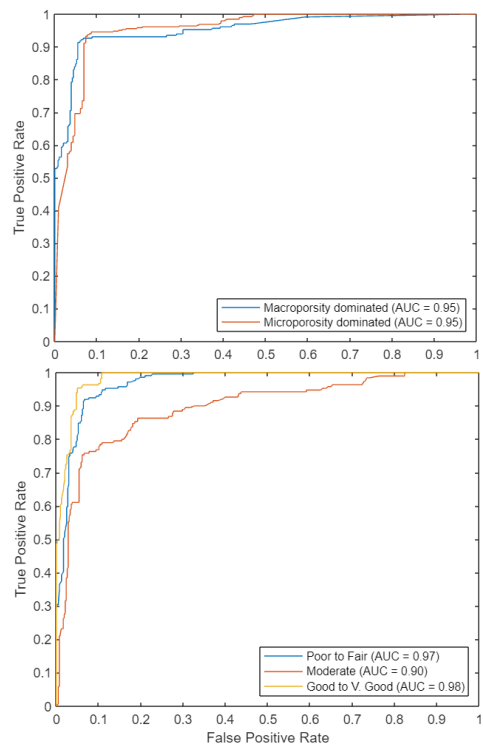

*Figure 7. ROC curves illustrating the performance of the Random Forest classifier for predicting permeability categories and microporosity domination. The curves depict the trade-off between the true positive rate and false positive rate at various threshold levels. AUC values, displayed in the legend, provide an overall measure of the model's accuracy in distinguishing between different classes.*

The model demonstrates strong predictive performance in estimating permeability and microporosity categories, as reflected by the high AUC values. For permeability prediction, AUC scores of 0.97, 0.90, and 0.98 for the (Poor to Fair), (Moderate), and (Good to Very Good) categories, respectively, highlight the model's ability to distinguish between different reservoir quality levels with high accuracy. Similarly, microporosity classification achieves an AUC of 0.95 for both macroporosity- and microporosity-dominated samples, reinforcing the reliability of the model in differentiating pore types. The consistently high AUC values suggest robust classification performance across varying thresholds, making the model a valuable tool for reservoir characterization and predictive analysis (Figure 7).

Table 3 shows how well the model predicted both permeability and microporosity domination for an independent holdout dataset. Most of the predictions align with the actual values, indicating the model's strong performance. For instance, samples C2-4 and C2-6 were correctly identified as (Poor to Fair) in terms of permeability, while samples like T3-2 and B6C-4 were accurately classified as (Good to V. Good).

When it comes to microporosity domination, the model also did well, correctly predicting samples such as C2-4, C2-6, and T9-1 as (Microporosity dominated). However, there were a few cases of misclassification. For example, sample B13-1 was predicted as (Good to V. Good) when it was actually (Moderate), and sample B6C-7, which was (Microporosity dominated), was incorrectly predicted as (Macroporosity dominated).

These misclassifications point out some of the model's limitations. It seems to struggle with certain subtleties in distinguishing between the classes, possibly due to complex overlaps in the dataset's features. Despite this, the majority of predictions were spot-on, showing that the model can be a reliable tool for assessing reservoir quality. It offers a solid foundation for future improvements, such as adding more detailed features or refining the model to better capture the nuances of microporosity and permeability.

Table 3. MICP Actual vs Predicted Values for the independent holdout dataset. Green cells show the correct predictions while red cells show wrong predictions.

| Sample | Actual Permeability | Predicted Permeability | Actual Porosity | Predicted Porosity |
|---|---|---|---|---|
| C2-4 | Poor to Fair | Poor to Fair | Microporosity dominated | Microporosity dominated |
| C2-6 | Poor to Fair | Poor to Fair | Microporosity dominated | Microporosity dominated |
| T3-2 | Good to V. Good | Good to V. Good | Macroporosity dominated | Macroporosity dominated |
| B6C-3 | Good to V. Good | Good to V. Good | Macroporosity dominated | Macroporosity dominated |
| B6C-4 | Good to V. Good | Good to V. Good | Macroporosity dominated | Macroporosity dominated |
| **B6C-7** | Good to V. Good | Good to V. Good | **Microporosity dominated** | **Macroporosity dominated** |
| **T8-2** | Poor to Fair | Poor to Fair | **Microporosity dominated** | **Macroporosity dominated** |
| T9-1 | Poor to Fair | Poor to Fair | Microporosity dominated | Microporosity dominated |
| B10-2 | Moderate | Moderate | Microporosity dominated | Microporosity dominated |
| B11-3 | Moderate | Moderate | Microporosity dominated | Microporosity dominated |
| **B13-1** | **Moderate** | **Good to V. Good** | Macroporosity dominated | Macroporosity dominated |

The uncertainty ranges played an important role in boosting the model's accuracy by incorporating the natural variability found in geological data. By training the model on a dataset that included this variability, it was exposed to a broader range of conditions, which helped to reduce the impact of measurement inconsistencies and prevent overfitting. This approach enabled the model to better adapt to new, unseen data. The inclusion of uncertainty analysis resulted in a significant reduction in prediction error, with approximately a 40% improvement for permeability and about a 30% improvement for microporosity. These findings underscore the importance of considering data variability in modeling, leading to more robust and reliable predictions, particularly in complex geological contexts. This demonstrates that incorporating such analysis is not merely an optional step but a crucial component in refining model performance for practical applications.

When compared to both simpler and more complex machine learning methods, the Random Forest model demonstrated superior performance in this study. Its ensemble learning approach provides greater stability and accuracy in predictions, which simpler models like linear regression or decision trees may struggle to achieve, particularly in the presence of non-linear geological data (Breiman, 2001). While more complex models like neural networks can sometimes offer higher accuracy, they often require significantly larger datasets and more computational resources. Random Forest strikes an effective balance, reducing the risk of overfitting while still handling

Uncertainty-Driven Modeling of Microporosity and Permeability in Clastic Reservoirs Using Random Forest<s>

heterogeneous datasets and missing data with robustness (Rodriguez-Galiano et al., 2015).

**4.3 Limitations and Future Improvements**

While the Random Forest model demonstrated strong performance, there are several limitations that should be addressed in future work. One key limitation is the model's sensitivity to the quality of input data. For example, misclassifications of microporosity domination in samples B6C-7 and T8-2 were likely due to the model's difficulty in handling laminated and interbedded sandstone facies. These facies were not directly included in the model as numerical features, which led to misclassifications. Incorporating more detailed sedimentological information into the model could improve its accuracy in these cases.

Sedimentological properties such as facies type have a significant impact on reservoir characteristics, but they are difficult to quantify numerically. Developing methods to integrate facies data into machine learning models, either through the use of numerical facies codes or by incorporating more detailed geological parameters, could enhance model accuracy and reduce misclassifications.

Future work could also explore the use of other machine learning algorithms, such as gradient boosting machines or neural networks, which may offer improvements in handling more complex relationships between variables. Additionally, increasing the diversity of the training dataset by including more geological formations and a wider range of sedimentological properties could further enhance the model's generalization capabilities.

The integration of uncertainty analysis with advanced machine learning techniques, combined with more detailed geological data, could lead to the development of even more accurate models for predicting reservoir quality. This would provide a valuable tool for both academic research and practical applications in the oil and gas industry.

## 5. Conclusion

The Random Forest machine learning model developed in this study successfully predicted permeability and microporosity domination (<2 μm) with high accuracy, achieving 90.91% for permeability and 81.82% for microporosity. By using key geological parameters such as total porosity, SGR data, and grain size distribution, the model proved to be a reliable tool for predicting reservoir quality in siliciclastic formations.

A major benefit of the model is its ability to predict complex properties like microporosity and permeability, which typically require expensive, time-consuming methods. Instead, the model leverages basic, cost-effective data such as porosity and grain size analysis, making it ideal for early-stage exploration and reducing both cost and uncertainty.

Incorporating uncertainty analysis enhanced the model's robustness, allowing it to generalize well across different formations. This approach is particularly useful in remote or offshore settings, where access to advanced laboratory testing is limited. The ability to generate accurate predictions using affordable data improves decision-making in exploration projects.

For future improvements, incorporating geological parameters like facies could increase prediction accuracy, and advanced imaging techniques such as backscattered SEM or microCT may provide more detailed insights into microporosity. Exploring different microporosity thresholds and expanding the dataset with more well logs could further enhance the model's performance and generalizability across different formations.

In conclusion, the Random Forest model, combined with uncertainty analysis, provides a cost-effective and powerful framework for predicting permeability and microporosity in clastic formations. By applying this model to similar geological contexts, reservoir quality predictions can be significantly improved, reducing uncertainty and optimizing exploration efforts, particularly when resources for expensive laboratory analyses are limited.

**Statements and Declarations**

**Author Contributions**

Conceptualization: MR; Literature Review: MR; Methodology: Lab Analysis MR; MR; Software Development and Validation: MR; Writing – original draft: MR; Writing – review & editing: MR, ME, and PL. The authors have read and approved the final manuscript.

**Funding**

This research was funded by the YUTP grant (015LC0-060).

**Data Availability Statement**

The data supporting the findings of this study are available on reasonable request from the corresponding author.

**Acknowledgments**

The author acknowledges the financial support from the Institute of Hydrocarbon Recovery, Universiti Teknologi PETRONAS.

**Competing Interests**

The author declares no relevant financial or non-financial competing interests that could have influenced the research findings.

**Ethical Compliance Statement**

Not applicable.

Uncertainty-Driven Modeling of Microporosity and Permeability in Clastic Reservoirs Using Random Forestmicrofocus X-ray computed tomography (μCT). *Geological Society, London, Special Publications*, *215*(1), 51–60.

Walter, S. D. (2005). The partial area under the summary ROC curve. *Statistics in Medicine*, *24*(13), 2025–2040. https://doi.org/10.1002/sim.2103

Wellmann, F., & Caumon, G. (2018). 3-D Structural geological models: Concepts, methods, and uncertainties. In *Advances in geophysics* (Vol. 59, pp. 1–121). Elsevier.

Wilson, M. D., & Pittman, E. D. (1977). Authigenic clays in sandstones; recognition and influence on reservoir properties and paleoenvironmental analysis. *Journal of Sedimentary Research*, *47*(1), 3–31.

Worden, R. H., & Burley, S. D. (2003). Sandstone Diagenesis: The Evolution of Sand to Stone. *Sandstone Diagenesis*, 1–44. https://doi.org/10.1002/9781444304459.ch

Xi, K., Cao, Y., Jahren, J., Zhu, R., Bjørlykke, K., Haile, B. G., Zheng, L., & Hellevang, H. (2015). Diagenesis and reservoir quality of the Lower Cretaceous Quantou Formation tight sandstones in the southern Songliao Basin, China. *Sedimentary Geology*, *330*, 90–107. https://doi.org/10.1016/j.sedgeo.2015.10.007

Xu, C., Fu, L., Lin, T., Li, W., & Ma, S. (2022). Machine learning in petrophysics: Advantages and limitations. *Artificial Intelligence in Geosciences*, *3*, 157–161. https://doi.org/10.1016/j.aiig.2022.11.004

Yong, H., Wenxiang, H., Yanli, Z., Bincheng, G., & Zhaopu, G. (2021). Uncertainty, sensitivity analysis and optimization of a reservoir geological model. *Marine Georesources & Geotechnology*, *39*(2), 129–139.

Zhong, Z., Carr, T. R., Wu, X., & Wang, G. (2019). Application of a convolutional neural network in permeability prediction: A case study in the Jacksonburg-Stringtown oil field, West Virginia, USA. *Geophysics*, *84*(6), B363–B373.

Zou, C., Zhao, L., Xu, M., Chen, Y., & Geng, J. (2021). Porosity prediction with uncertainty quantification from multiple seismic attributes using random forest. *Journal of Geophysical Research: Solid Earth*, *126*(7), e2021JB021826.
13